\documentclass[reprint,
superscriptaddress,
nobibnotes,
amsmath,amssymb,
aps,
longbibliography
]{revtex4-2}

\usepackage{graphicx}
\usepackage{dcolumn}
\usepackage{bm}
\usepackage{braket}
\usepackage{hyperref}
\usepackage{changes}
\usepackage{cancel}
\hypersetup{
     colorlinks   = true,
     linkcolor    = blue,
     citecolor    = blue,
}

\begin{document}

\title{Comparison of the canonical transformation and energy functional formalisms\\ for \textit{ab initio} calculations of self-localized polarons}


\author{Yao Luo}%
\affiliation{Department of Applied Physics and Materials Science, California Institute of Technology, Pasadena, California 91125, USA}

\author{Benjamin K. Chang}%
\affiliation{Department of Applied Physics and Materials Science, California Institute of Technology, Pasadena, California 91125, USA}


\author{Marco Bernardi}
\affiliation{Department of Applied Physics and Materials Science, California Institute of Technology, Pasadena, California 91125, USA}
\email{bmarco@caltech.edu}

\begin{abstract}
\noindent
In materials with strong electron-phonon (e-ph) interactions, charge carriers can distort the surrounding lattice and become trapped, forming self-localized (small) polarons.
We recently developed an \textit{ab initio} approach based on canonical transformations to efficiently compute the formation and energetics of small polarons~\cite{Lee2021}.
A different approach based on a Landau-Pekar energy functional has been proposed in the recent literature~\cite{Sio201901, Sio201902}.
In this work, we analyze and compare these two methods in detail. We show that the small polaron energy is identical in the two formalisms when using the same polaron wave function. We also show that our canonical transformation formalism can predict polaron band structures and can properly treat zero- and finite-temperature lattice vibration effects, although at present using a fixed polaron wave function. Conversely, the energy functional approach can compute the polaron wave function, but as we show here it neglects lattice vibrations and cannot address polaron self-localization and thermal band narrowing.
%
Taken together, this work relates two different methods developed recently to study polarons from first-principles, highlighting their merits and shortcomings and discussing them both in a unified formalism.
\end{abstract}

\maketitle
%
%
\section{\label{sec1:intro}Introduction}
\vspace{-10pt}
Self-localized (small) polarons are charge carriers that interact strongly with the lattice vibrations, becoming trapped as a result of their local lattice distortion~\cite{Emin1982}. Small polarons are key to understanding the physical properties of materials with strong or localized electron-phonon ($e$-ph) interactions, including alkali halides, organic molecular crystals, transition metal oxides, and some glasses~\cite{Emin1982}.
Small polarons can be studied experimentally using spectroscopy, diffraction, and microscopy techniques sensitive to the local lattice distortion~\cite{Goovaerts1978, Sezen2015, Pastor2019}. Their signatures are also found in transport properties: as small polarons move only in response to certain vibrations of the surrounding atoms, they are associated with a low charge carrier mobility ($<1$ cm$^2$/Vs), which typically increases with temperature due to thermally activated small-polaron hopping~\cite{Nagels1963, Crevecoeur1970}. As a result, small polarons are detrimental in many technological applications where electrical transport limits device efficiency. 
\\ 
\indent
%
%
Theoretical treatments of small polarons span a wide range of analytic and numerical techniques~\cite{Alexandrov2010, Emin-book}. Focusing on first-principles approaches based on density functional theory (DFT), a key goal has been the development of parameter-free, quantitative predictions of the energetics and dynamics of small polarons.
The conventional approach employs DFT calculations on supercells with excess charge or defects added explicitly~\cite{Varley2012, Kokott2018, Reticcioli2019b, Pasquarello2020}.
Yet, recent work has developed a different family of first-principles methods aimed at computing polarons using only a unit cell of the material with \textit{ab initio} $e$-ph calculations~\cite{Lee2021,Sio201901,Sio201902}. The goal of these approaches is two-fold: reducing computational cost by avoiding calculations on large supercells with many atoms, and formulating rigorous polaron calculations by combining many-body techniques with first-principles theories.\vspace{10pt}
\\
\indent

Within this recent body of work, we formulated an efficient approach based on the canonical transformation formalism~\cite{Lee2021}; this method can compute the small-polaron formation energy in a localized Wannier basis starting from a trial polaron wave function. Its computational cost is a minimal overhead to a DFT calculation on a unit cell, enabling investigations of small polarons in a wide range of materials with minimal computational effort~\cite{Lee2021}.
%
A different approach, proposed by Sio et al.~\cite{Sio201901,Sio201902}, uses a Landau-Pekar-type energy functional to obtain coupled equations for the polaron wave function and its associated atomic displacements. Solving these equations on a fine reciprocal-space grid can provide the polaron formation energy and wave function.
\\
\indent
Here we compare in detail these two methods, and explain how they address various aspects of polaron physics. 
We first discuss the canonical transformation approach, showing example calculations of the polaron energy and band structure in an ionic insulator (NaCl) and an organic semiconductor (naphthalene). 
%
We show that for small polarons the canonical transformation and energy functional formalisms give the same polaron energy when using the same polaron wave function.
We also present a generalization of the canonical transformation formalism that can compute the polaron wave function and treat both small and delocalized polarons. This formulation allows us to relate the canonical transformation and energy functional approaches, and discuss differences in how they treat the polaron wave function, thermal effects, band narrowing and polaron self-localization. Of the two approaches, we show that only the canonical transformation framework can guarantee polaron self-localization (due to vanishing hopping) and correctly treat lattice vibrations and thermal effects.
%
%
%
%
Taken together, our work advances the formulation of rigorous methods to study polarons with first-principles calculations on a unit cell of the material.
%
\section{Canonical Transformation}
%
Inspired by analytic treatments of small polarons~\cite{Holstein1959, Mahan2000}, we derive an effective polaron Hamiltonian by transforming the coordinates to the distorted lattice configuration induced by the localized charge carrier. This technique was introduced by Lee, Low, and Pines to study the large polaron problem~\cite{Lee1953}, and it can be traced back to a method used by Tomonaga \cite{Tomonaga1947} to solve a meson problem. The transformation is closely related to the one that diagonalizes the charged harmonic oscillator (CHO) in an external electric field \cite{Mahan2000}. Therefore, we first briefly review the CHO treatment to set the stage for the polaron canonical transformation.\\
\subsection{Analogy with the charged harmonic oscillator}
\label{sec:toy_model}
The Hamiltonian of a one-dimensional CHO is
\begin{align}
H^{\textrm{(CHO)}} = \frac{1}{2m} p^2 + \frac{1}{2} m \omega^2 x^2 + eEx\,, \nonumber
\end{align}
where $x$ is the position and $p$ the momentum of a particle with mass $m$ and charge $e$. The oscillator frequency is $\omega$, and $E$ is the external electric field that couples to the system (here and below, we set $\hbar=1$). The CHO Hamiltonian can be solved by completing the square and shifting to a new coordinate system:
\begin{gather}
\label{coordinate_shift}
x^{\prime}=x-x_0,\\
\label{H_CHO_x_space}
{H^{\prime}}^{\textrm{(CHO)}} = \frac{1}{2m} p^2 + \frac{1}{2} m \omega^2 {x^{\prime}}^2
                              - \frac{e^2 E^2}{2 m \omega^2}.
\end{gather}
The particle then oscillates around the new equilibrium point, $x_0=-eE/m\omega^2$, with the same frequency $\omega$ as in the absence of the external field.\\
\indent
We study this transformation by introducing creation and annihilation operators for the harmonic oscillator,
\begin{align}
x&=\frac{1}{\sqrt{2m\omega}} (b^{\dagger}+b)\nonumber\\
p&=i \sqrt{\frac{m\omega}{2}}(b^{\dagger}-b)\nonumber.
\end{align}
%
The CHO Hamiltonian becomes
\begin{align}
\label{H_CHO}
H^{\textrm{(CHO)}} = \omega ( b^{\dagger} b + \frac{1}{2} )
      + \omega g ( b + b^{\dagger} ),
\end{align}
with the coupling constant $g=eE/\sqrt{2m\omega^3}$. As implied by Eq.~(\ref{H_CHO_x_space}), this Hamiltonian can be solved exactly by a translation to the new equilibrium position $x_0$, as can be achieved with the canonical transformation of operators $\mathcal{O} \rightarrow \widetilde{\mathcal{O}} = e^S \mathcal{O}e^{-S}$, with the translation generated by
\begin{align}
\label{S_CHO}
S^{\textrm{(CHO)}}= ipx_0 =  g(b^{\dagger}-b).
\end{align}
Using the Baker-Campbell-Hausdorff (BCH) formula
\begin{align}
\label{BCH_formula}
\widetilde{\mathcal{O}}=
e^{S} \mathcal{O} e^{-S} =
    \mathcal{O} + [S, \mathcal{O}] + \frac{1}{2!}[S,[S,\mathcal{O}]]
    + \cdots,
\end{align}
we obtain the \textit{transformed} annihilation operator and Hamiltonian (denoted by the tilde symbol):
\begin{gather}
\label{b_tilde_CHO}
\widetilde{b}=b-g,\\
\widetilde{H}^{\textrm{(CHO)}}=\omega(b^{\dagger}b+\frac{1}{2}) - \omega g^2.
\label{transformed_H_CHO}
\end{gather}
\indent
The second term in $\widetilde{H}^{\textrm{(CHO)}}$ is the decrease in potential energy associated with stretching the oscillator spring:
\begin{gather}
\label{CHO_energy_decrease}
- \omega g^2=\frac{1}{2}m \omega^2 {x_0}^2 + eEx_0.
\end{gather}
The meaning of the operators $b$ and $b^{\dagger}$ after the  transformation can be understood by analyzing the transformed position operator
\begin{align}
\label{coordinate_shift_2}
\widetilde{x}=x+[S^{\textrm{(CHO)}}, x]=x+x_0
\end{align}
where the $x$ coordinate on the right-hand side now measures the position relative to $x_0$; this way, the coordinates $x$ and $\widetilde{x}$ in Eq.~(\ref{coordinate_shift_2}) correspond to $x^{\prime}$ and $x$ in Eq.~(\ref{coordinate_shift}), respectively. Expressed in terms of creation and annihilation operators, this transformed coordinate becomes
\begin{align}
\label{CHO_x_tilde}
\widetilde{x}=\frac{1}{\sqrt{2m\omega}}(\widetilde{b}^{\dagger}+\widetilde{b})=\frac{1}{\sqrt{2m\omega}}(b^{\dagger}+b) + x_0.
\end{align}
Therefore, by comparison it is clear that $b^{\dagger}$ in $\widetilde{H}^{\textrm{(CHO)}}$ creates oscillation quanta relative to the new equilibrium point $x_0$, as illustrated in Fig.~\ref{fig:schematics}(a)-(b).
\\
\indent
%
%
The ground state wave function of the CHO, denoted here as $|0\rangle$, is centered at $x_0$, and is annihilated by the operator $b$, so that $b|0\rangle=0$. Denote as $|\widetilde{0}\rangle$ the original ground state (before applying the electric field) which is annihilated by $\widetilde{b}$. Then, since
\begin{align}
b|\widetilde{0}\rangle = ( \widetilde{b} + g )|\widetilde{0}\rangle = g |\widetilde{0}\rangle \nonumber
\end{align}
we can see that $|\widetilde{0}\rangle$ is a coherent state of $b$ with eigenvalue $g$, and therefore we have
\begin{align}
\label{displacement_operator_0}
|\widetilde{0}\rangle= \textrm{exp}\left[{ g (b^{\dagger}-b) }\right] |0\rangle = \textrm{exp}\left[ S^{\textrm{(CHO)}} \right] |0\rangle.
\end{align}
The exponential factor connecting $|\widetilde{0}\rangle$ and $|0\rangle$ is the translation operator in Eq.~(\ref{S_CHO}). This result shows that the ground states before and after applying the electric field are related via a translation by $x_0$~\cite{Mahan2000}.\\
\vspace{20pt}
%
%
%
%
%
\newpage
\subsection{Derivation of the polaron Hamiltonian}
\vspace{-10pt}
We now carry out a transformation analogous to the CHO case to obtain an effective polaron Hamiltonian. The starting point is the $e$-ph Hamiltonian in the electronic Wannier and phonon momentum basis~\cite{Perturbo},
\begin{align}
\label{electron_phonon_hamiltonian}
H = & \sum_{mn} \varepsilon_{mn} a^{\dagger}_{m}a_{n}
      + \sum_{\textbf{Q}} \omega_{\textbf{Q}}
      \left(b^{\dagger}_{\textbf{Q}}b_{\textbf{Q}}
      + \frac{1}{2}\right)\\
    & +  \frac{1}{ \sqrt{N_{\Omega}} }
      \sum_{mn}\sum_{\textbf{Q}}\omega_{\textbf{Q}}
      g_{\textbf{Q}mn}
      \left(b^{\dagger}_{\textbf{Q}}+b_{-\textbf{Q}}\right)
      a^{\dagger}_{m}a_{n}\nonumber,
\end{align}
where $n=j_n \textbf{R}_{n}$ is a collective index labelling the $j_n$-th Wannier function (WF) in the unit cell with origin at the Bravais lattice vector $\textbf{R}_{n}$, and $a_n = a_{j_n \textbf{R}_{n}}$ is the corresponding electron annihilation operator; $b_{\textbf{Q}}$ is the phonon annihilation operator, where $\textbf{Q}$ is a collective label for the phonon mode $\nu$ and momentum $\textbf{q}$.
The hopping strength and phonon energy are denoted as $\varepsilon_{mn}$ and $\omega_{\textbf{Q}}$, respectively, and $N_{\Omega}$ is the number of unit cells in the crystal.
\\
\indent
The $e$-ph coupling matrix elements in the Wannier basis, denoted as $g_{\textbf{Q}mn}$, are unitless and do not include the phonon frequency factor, different from the standard convention \cite{Perturbo}.
They are obtained by transforming unitless $e$-ph matrix element in momentum space, $\tilde{g}_{ij\nu} (\mathbf{k},\mathbf{q})=  g_{ij\nu} (\mathbf{k},\mathbf{q}) / (\hbar \omega_{\nu \mathbf{q}})$, to the electron Wannier basis, where $g_{ij\nu} (\mathbf{k},\mathbf{q})$ are defined in Eq.~(24) of Ref.~\cite{Perturbo}, and $i$ and $j$ are band indices.
Using the notation in Ref.~\cite{Perturbo}, the explicit definition is
$g_{\textbf{Q}mn} \equiv \tilde{g}_{mn\nu}(\mathbf{R}_e,\mathbf{q}) =
\frac{1}{N_e} \sum_{\mathbf{k},ij} e^{-i \mathbf{k}\cdot \mathbf{R}_e} \mathcal{U}^\dagger_{mi} (\mathbf{k} + \mathbf{q})
\tilde{g}_{ij\nu} (\mathbf{k},\mathbf{q}) \mathcal{U}_{jn}(\mathbf{k})$,
where $\mathcal{U}$ are unitary Wannier matrices~\cite{Perturbo}. Also recall that the $e$-ph coupling needs to satisfy the relation $g^{*}_{\textbf{Q}mn}= g_{-\textbf{Q}nm}$ for the Hamiltonian to be Hermitian.\\
\indent
The $e$-ph interaction term in Eq.~(\ref{electron_phonon_hamiltonian}) has the same form as in $H^{\textrm{(CHO)}}$ in Eq.~(\ref{H_CHO}), but now the external field coupling to the ``spring'' of each phonon mode is controlled by the electronic configuration through the factor $a^{\dagger}_ma_n$. In analogy with Eq.~(\ref{S_CHO}), we stretch the spring of each phonon mode to a new equilibrium position using the canonical transformation $\mathcal{O} \rightarrow \widetilde{\mathcal{O}} = e^S \mathcal{O}e^{-S}$, with the generator defined as
\begin{gather}
\label{polaron_S}
S = \sum_{mn}C_{mn}a^{\dagger}_{m}a_{n},\\
\label{polaron_C_mn}
C_{mn} =  \frac{1}{ \sqrt{N_{\Omega}} } \sum_{\textbf{Q}} B_{\textbf{Q}mn}
    (b^{\dagger}_{\textbf{Q}}-b_{-\textbf{Q}}).
\end{gather}
Above, we introduced the undetermined distortion coefficients $B_{\textbf{Q}mn}$ which, analogous to the coupling $g$ in the CHO example, quantify the stretching of the spring associated with each phonon mode, as we show below.
To make the transformation unitary, we impose the conjugate relation $B^{*}_{\textbf{Q}mn}= B_{-\textbf{Q}nm}$, so that the operator $S$ is anti-Hermitian.\vspace{10pt}
\\

\indent
To obtain the transformed electron and phonon operators, we  compute their commutators with $S$:
\begin{align}
[S,a_{m}] &= - \sum_{n}C_{mn}a_{n}\nonumber,\\
[S,b_{\textbf{Q}}] &=
    - \frac{1}{ \sqrt{N_{\Omega}} } \sum_{mn}B_{\textbf{Q}mn}a^{\dagger}_{m}a_n\nonumber.
\end{align}
Then using the BCH formula in Eq.~(\ref{BCH_formula}), we have
\begin{align}
\label{a_tilde}
\widetilde{a}_m &= \sum_{n} e^{-C}_{mn}a_n,\\
\label{b_tilde}
\widetilde{b}_{\textbf{Q}} &= b_{\textbf{Q}}
    - \frac{1}{ \sqrt{N_{\Omega}} } \sum_{mn}B_{\textbf{Q}mn}a^{\dagger}_{m}a_n,
\end{align}
where $e^{-C}_{mn}$ is a shorthand notation for the phonon operator
\begin{align}
\label{exp_C_mn}
e^{-C}_{mn} =
    \delta_{mn} - C_{mn}
    + \frac{1}{2!}\sum_{i}C_{mi}C_{in}-\cdots.
\end{align}
%
%
%
\begin{figure}[!t]
\includegraphics[width=0.97\linewidth]{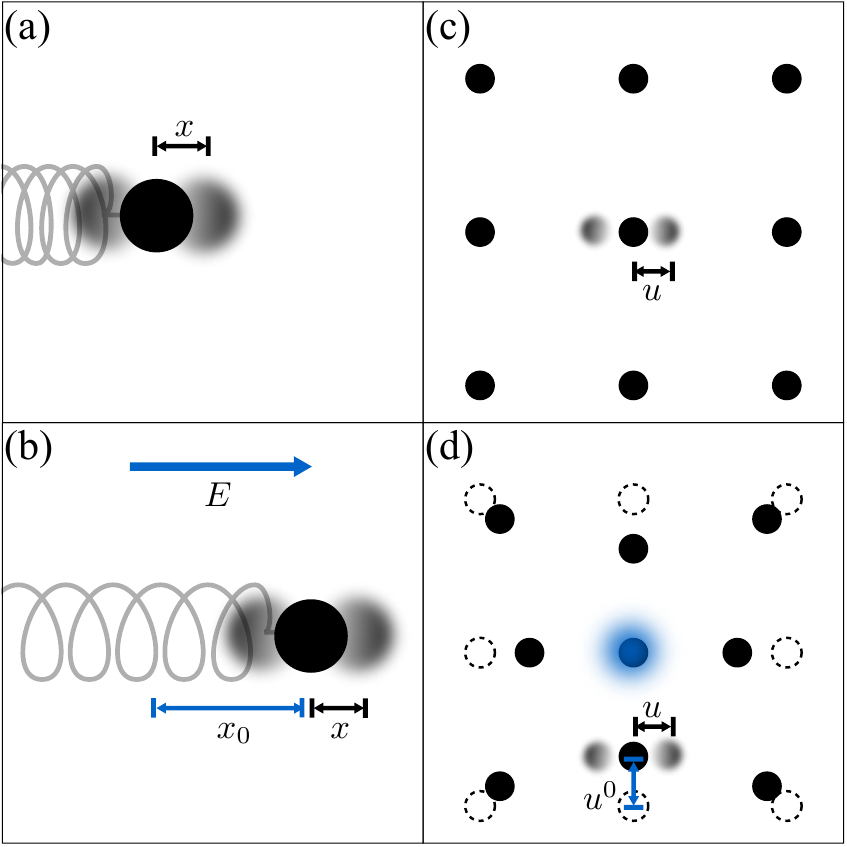}
\caption{Schematic of the charged harmonic oscillator, (a) without and (b) with the electric field applied. By analogy, panel (c) shows the unperturbed crystal lattice, and panel (d) the distorted lattice in the presence of the excess charge carrier. In panel (b), the spring is stretched by an amount of $x_0$; analogously, in panel (d) the static lattice distortion is defined as $u^0$. The coordinates of the oscillation quanta relative to the new (distorted) equilibrium positions are denoted as $x$ or $u$, in (b) and (d) respectively.} 
\label{fig:schematics}
\end{figure}
\noindent
The physical intuition is that the distortion coefficients $B_{\textbf{Q}mn}$ quantify how the transformation stretches each phonon spring, as is seen by comparing Eq.~(\ref{b_tilde}) with the CHO case in Eq.~(\ref{b_tilde_CHO}).
\\

\indent
It is instructive to examine the lattice displacements in real space~\cite{Bernardi-review}. Denoting as $u_{cs\alpha}$ the displacement of the atom $s$ in unit cell $c$ along the Cartesian coordinate $\alpha$, the atomic positions in the transformed basis become:
\begin{align}
\label{utilde}
\widetilde{u}_{cs\alpha} = u_{cs\alpha} + u^{0}_{cs\alpha}\nonumber
= \sum_{\textbf{Q}} \frac{1}{\sqrt{2M_{s}\omega_{\textbf{Q}}N_{\Omega}}}
        e^{s\alpha}_{\textbf{Q}}e^{i\textbf{q}\cdot \textbf{R}_{c}} \times \nonumber\\
\left[\left(b^{\dagger}_{\textbf{Q}}+b_{-\textbf{Q}}\right)
        -\frac{2}{\sqrt{N_{\Omega}}} \sum_{mn}
        B_{-\textbf{Q}mn}
        a^{\dagger}_m a_n \right], 
\end{align}
in full analogy with Eq.~(\ref{CHO_x_tilde}) for the CHO (above, $M_s$ is the mass of atom $s$ and $e^{s\alpha}_{\textbf{Q}}$ is the phonon eigenvector).
From Eq.~(\ref{utilde}), we see that the lattice distortion induced by the localized polaron gives new equilibrium atomic positions,
\begin{align}
\label{lattice_distortion}
u^0_{cs\alpha} = - \frac{1}{N_{\Omega}} \sum_{\textbf{Q}} \sqrt{\frac{2}{M_s \omega_{\textbf{Q}}}}
        e^{s\alpha}_{\textbf{Q}}e^{i\textbf{q}\cdot \textbf{R}_{c}}
        \sum_{mn} B_{-\textbf{Q}mn} a^{\dagger}_m a_n .
\end{align}
%
%
\noindent
The lattice distortion depends on the polaron electronic state, as is clear from the factors $a^{\dagger}_m a_n$. In addition, the distortion coefficients $B_{\textbf{Q}mn}$ can now be identified as generalized Fourier transforms of the lattice distortion. Figure~\ref{fig:schematics}(c)-(d) illustrate the lattice distortion to new equilibrium positions, displaced by $\mathbf{u}^0_{cs}$ relative to the pristine lattice, together with the vibrational coordinates $\mathbf{u}_{cs}$ relative to the distorted lattice.
\\
\indent
After the transformation, the operators $a^{\dagger}_n$ and $b^{\dagger}_\textbf{Q}$ create polarons and phonons in the distorted lattice, respectively, analogous to the CHO. The factor $e^{-C}_{mn}$ in Eq.~(\ref{a_tilde}) is analogous to the translation operator $\textrm{exp}[-S^{(\textrm{CHO})}]$, and thus it relaxes the distortion and sends each phonon to their corresponding vibrational mode of the undistorted lattice.
Loosely speaking, Eq.~(\ref{a_tilde}) implies that annihilating a polaron from a given site amounts to first annihilating the electronic state from the distorted lattice and then removing the lattice distortion.\\
\indent
The transformed Hamiltonian is derived by replacing the electron and phonon operators with their transformed counterparts. We obtain:
\begin{align}
\widetilde{H}=& \sum_{mn}\widetilde{\varepsilon}_{mn}a^{\dagger}_{m}a_n
          + \sum_{\textbf{Q}}\omega_{\textbf{Q}}
             (b^{\dagger}_{\textbf{Q}}b_{\textbf{Q}}+\frac{1}{2})
             \nonumber\\
          &+ \frac{1}{ \sqrt{N_{\Omega}} }
             \sum_{mn\textbf{Q}}\omega_{\textbf{Q}}
             (\widetilde{g}_{\textbf{Q}mn}-B_{\textbf{Q}mn})
             (b^{\dagger}_{\textbf{Q}} + b_{-\textbf{Q}})
             a^{\dagger}_{m}a_n
             \nonumber\\
          &+ \frac{1}{ N_{\Omega} }
             \sum_{mnij\textbf{Q}}\omega_{\textbf{Q}}
             B_{-\textbf{Q}ij}(B_{\textbf{Q}mn}-2\widetilde{g}_{\textbf{Q}mn})
             a^{\dagger}_{i}a_{j}
             a^{\dagger}_{m}a_{n},\nonumber
\end{align}
where the transformed hopping and $e$-ph coupling matrices $\widetilde{\varepsilon}_{mn}$ and $\widetilde{g}_{\textbf{Q}mn}$, denoted collectively as $\widetilde{M}_{mn}$, are defined as
\begin{align}
\label{M_tilde}
\widetilde{M}_{mn} = \sum_{ij}e^{C}_{mi}M_{ij}e^{-C}_{jn}.
\end{align}
%
%
%
These transformed matrices encode the effect of a polaron dragging the lattice distortion. For example, the hopping term in the transformed Hamiltonian, $\sum_{mn} a^{\dagger}_m \widetilde{\varepsilon}_{mn} a_n = \sum_{mnij} a^{\dagger}_m \, ( e^{C}_{mi}\varepsilon_{ij} e^{-C}_{jn})\, a_n$, shows that if a polaron hops to a nearby site, the lattice distortion is first removed at the original site by $e^{-C}_{jn}$, then the electron hops with amplitude $\varepsilon_{ij}$, and then the distortion is created at the new site by $e^{C}_{mi}$.
\\
\indent
Note that the transformed matrices $ \widetilde{M}$ in Eq.~(\ref{M_tilde}) still contain phonon operators $b_{\textbf{Q}}$ in the $e^{\pm C}$ terms. Following Holstein~\cite{Holstein1959}, we take the thermal average of the phonon operators, obtaining averaged transformed matrices $\langle \widetilde{M}\rangle_{mn}$ as explained below.
After collecting terms, and using the identity
\begin{align}
a^{\dagger}_{i}a_{j}a^{\dagger}_{m}a_{n} =
    a^{\dagger}_{i}a_{n}\delta_{mj}
    -a^{\dagger}_{i}a^{\dagger}_{m}a_{j}a_{n}
    \nonumber,
\end{align}
we obtain our effective polaron Hamiltonian:

  \begin{align}
\label{H_tilde_ta}
\widetilde{H}=& \sum_{mn} E_{mn} a^{\dagger}_{m} a_n
          + \sum_{\textbf{Q}} \omega_{\textbf{Q}}
             (b^{\dagger}_{\textbf{Q}}b_{\textbf{Q}}+\frac{1}{2})
             \\
          &+ \frac{1}{ \sqrt{N_{\Omega}} }
             \sum_{mn\textbf{Q}}\omega_{\textbf{Q}}
             G_{\textbf{Q}mn}
             (b^{\dagger}_{\textbf{Q}} + b_{-\textbf{Q}})
             a^{\dagger}_{m}a_n
             \nonumber\\
          &- \frac{1}{ N_{\Omega} }
             \sum_{ijmn\textbf{Q}}
             V_{\textbf{Q}ijmn}
             a^{\dagger}_{i} a^{\dagger}_{m} a_{j} a_{n}\,,\nonumber
\end{align}
where the polaron hopping strength $E_{mn}$, the residual polaron-phonon (pl-ph) coupling constant $G_{\textbf{Q}mn}$, and the effective polaron-polaron (pl-pl) interaction $V_{\textbf{Q}ijmn}$ are defined respectively as
\begin{gather}
E_{mn} = \langle \widetilde{\varepsilon}\rangle_{mn}  +
     \frac{1}{N_{\Omega}}\sum_{i\textbf{Q}} \omega_{\textbf{Q}} B_{-\textbf{Q}mi}
     \left( B_{\textbf{Q}in} - 2 \langle \widetilde{g}_{\textbf{Q}} \rangle_{in} \right),
     \nonumber\\
\label{polaron_parameters}
G_{\textbf{Q}mn} = \langle \widetilde{g}_{\textbf{Q}} \rangle_{mn} - B_{\textbf{Q}mn},
     \vphantom{\sum_{\textbf{Q}}}\\
V_{\textbf{Q}ijmn} = \omega_{\textbf{Q}}
     B_{-\textbf{Q}ij} \left( B_{\textbf{Q}mn} - 2 \langle \widetilde{g}_{\textbf{Q}} \rangle_{mn} \right).
     \vphantom{\sum_{\textbf{Q}}}\nonumber
\end{gather}
\indent
Due to the thermal averaging process, all these quantities are now c-numbers rather than phonon operators. Therefore the Hamiltonian in Eq.~(\ref{H_tilde_ta}) reduces to an effective tight-binding model, which can be studied with standard approaches. In this work, we assume that the carrier concentration is low enough to neglect the pl-pl interaction $V_{\textbf{Q}ijmn}$.\\
\subsection{Thermal average}
\label{sec:thermal_average}
\vspace{-10pt}
The thermal average of the transformed matrices, $\langle \widetilde{M}\rangle_{mn} $, appears above in the polaron Hamiltonian and needs to be evaluated. This thermal average admits an exact expression only in the Holstein model~\cite{Mahan2000}, in which all $e$-ph coupling constants $g_{\textbf{Q}mn}$ are zero unless $m\!=\!n$.
However, in the general case, a closed-form expression for $\langle \widetilde{M}\rangle_{mn}$ cannot be derived without assuming that the distortion coefficients commute with each other: $[B_{\textbf{Q}}, B_{\textbf{Q}^{\prime}}]_{mn} = 0$ for all pairs of $\textbf{Q}$ and $\textbf{Q}^{\prime}$~\cite{Hannewald200401}. There are two main approaches to calculate the thermal average in the general case, the first uses the Feynman disentangling of operators \cite{Munn1985} and the second the BCH formula \cite{Hannewald200401}. Here we follow the latter strategy and derive the expression for the thermally-averaged transformed matrices.\\
\indent
First, using the BCH formula in Eq.~(\ref{BCH_formula}), we have
\begin{align}
\langle \widetilde{M}\rangle_{mn}  =
        \langle e^{C}M e^{-C}\rangle _{mn}
    = \langle M + \frac{1}{2!}[C,[C,M]]
        + \cdots \rangle _{mn}\nonumber,
\end{align}
where the angle brackets $\langle\cdots\rangle$ indicate a thermal average over phonon states. In this expression, terms with an odd number of $C_{mn}$ operators vanish because the thermal average of an odd number of $b_{\textbf{Q}}$ or $b^{\dagger}_{\textbf{Q}}$ is zero. Substituting the definition of $C_{mn}$, we get
\begin{align}
\langle \widetilde{M}\rangle_{mn}  =  &
        \langle  \sum_{i}  \frac{1}{(2i)!}
        [C,[\cdots[C,[C,M]]\cdots] \rangle_{mn}
        \nonumber\\
    =  & \sum_{i}  \frac{1}{(2i)! N_{\Omega}^i}
        [\sum_{\textbf{Q}_1}B_{\textbf{Q}_1},[\cdots
        [\sum_{\textbf{Q}_{2i}}B_{\textbf{Q}_{2i}},
            M]\cdots]]\times\nonumber\\
     & \hphantom{\sum_{i}\sum_{i}B}
        \langle(b^{\dagger}_{\textbf{Q}_1}-b_{-\textbf{Q}_1})\cdots
        (b^{\dagger}_{\textbf{Q}_{2i}}-b_{-\textbf{Q}_{2i}})
        \rangle_{mn}\nonumber,
\end{align}
where the phonon operator part can be factored out in the last equality because all permutations of $(b^{\dagger}_{\textbf{Q}}-b_{-\textbf{Q}})$ give the same thermal average. Next, we apply the Wick theorem and use the well-known thermal averages $\langle b^{\dagger}_{\textbf{Q}} b_{\textbf{Q}} \rangle = N_{\textbf{Q}}$ and $\langle b_{\textbf{Q}} b^{\dagger}_{\textbf{Q}} \rangle = N_{\textbf{Q}}+1$, where $N_{\textbf{Q}}$ is the phonon thermal occupation, obtaining
\begin{align}
\langle \widetilde{M}\rangle_{mn}
    =  \sum_{i} \frac{(-1)^{i}}{(2i)!N_{\Omega}^i}
       \sum_{\textbf{Q}_1 \cdots \textbf{Q}_{i}}
       (2N_{\textbf{Q}_1}+1)\cdots(2N_{\textbf{Q}_i}+1)
       \times\nonumber\\
\sum_{\textrm{all pairings}}
        [B_{\textbf{Q}_1},[B_{-\textbf{Q}_1},[\cdots
        [B_{\textbf{Q}_i},[B_{-\textbf{Q}_i},M]]
        \cdots]]]_{mn}.\nonumber
\end{align}
Assuming all the distortion coefficients $B_{\textbf{Q}}$ commute with each other, the commutator factors are identical for every possible pairing. Under this assumption, each of the $(2i)!/2^i i!$ possible pairings gives the same contribution, and thus
\begin{align}
\langle \widetilde{M}\rangle_{mn}
    =  \sum_{i} \frac{(-1)^{i}}{N_{\Omega}^i i!}
       \sum_{\textbf{Q}_1 \cdots \textbf{Q}_{i}}
       (N_{\textbf{Q}_1}+\frac{1}{2})\cdots(N_{\textbf{Q}_i}+\frac{1}{2})
       \times\nonumber\\
[B_{\textbf{Q}_1},[B_{-\textbf{Q}_1},[\cdots
        [B_{\textbf{Q}_i},[B_{-\textbf{Q}_i},M]]
        \cdots]]]_{mn}.\nonumber
\end{align}
Defining the linear operator $\Lambda$ on $M$ as
\begin{align}
\label{def_Lambda}
\sum_{\alpha \gamma} \Lambda_{mn\textrm{,}\alpha \gamma}  M_{\alpha \gamma}
= \frac{1}{N_{\Omega}}\sum_{\textbf{Q}}
        ( N_{\textbf{Q}} + \frac{1}{2} )
        \left[ B_{\textbf{Q}},\left[B_{-\textbf{Q}},M \right]  \right]_{mn},
\end{align}
we derive the final expression for the thermal averages
\begin{align}
\label{thermal_avg_M_tilde}
\langle \widetilde{M}\rangle_{mn}
    =  \sum_{\alpha \gamma} e^{-\Lambda}_{mn\textrm{,}\alpha \gamma}  M_{\alpha \gamma}.
\end{align}
\indent
In the special case where all the nonlocal distortion coefficients vanish, i.e. if $B_{\textbf{Q}mn}$ is nonzero only if $m= n$, the expressions in Eqs.~(\ref{def_Lambda}) and (\ref{thermal_avg_M_tilde}) are exact~\cite{Mahan2000, Munn1980, Hannewald200401}.
In this case, using $B_{\textbf{Q}mn}=B_{\textbf{Q}mm}\delta_{mn}$ in Eq.~(\ref{def_Lambda}), we obtain:
\begin{gather}
\Lambda_{mn\textrm{,}\alpha \gamma} = \lambda_{mn} \delta_{m\alpha}\delta_{n\gamma},
        \vphantom{\frac{1}{N_{\Omega}}}\nonumber\\
\label{thermal_average_lambda_local}
\lambda_{mn} (T) = \frac{1}{N_{\Omega}}\sum_{\textbf{Q}}
        \left[ N_{\textbf{Q}}(T) + \frac{1}{2} \right]
        \big| B_{\textbf{Q}mm} - B_{\textbf{Q}nn} \big|^2.
\end{gather}
With this definition, the thermal average of the transformed matrix becomes
\begin{align}
\label{thermal_average_M_tilde_local}
\langle \widetilde{M}\rangle_{mn} = \textrm{exp}\left( -\lambda_{mn} \right) M_{mn}.
\end{align}
%
Under these assumptions, the explicit expressions for the transformed polaron hopping and $e$-ph coupling are $\langle \widetilde{\varepsilon}\rangle_{mn} \!=\! \textrm{exp}\left( -\lambda_{mn} \right) \varepsilon_{mn}$ and $\langle \widetilde{g}_{\textbf{Q}}\rangle_{mn} \!=\! \textrm{exp}\left( -\lambda_{mn} \right) g_{\textbf{Q}mn}$, respectively.
\\
\indent
These expressions greatly simplify the evaluation of $\langle \widetilde{\varepsilon}\rangle_{mn}$ and $\langle \widetilde{g}_{\textbf{Q}}\rangle_{mn}$, which now involve only the exponential of a specific matrix element, rather than an exponential of an entire matrix as in Eq.~(\ref{thermal_avg_M_tilde}). Below, we refer to $e^{-\lambda_{mn}}$ as the \textit{band narrowing factor} because the polaron hopping amplitude $\langle \widetilde{\varepsilon}\rangle_{mn}$ is suppressed by $e^{-\lambda_{mn}}$.
%
%
%
%
\subsection{Small polaron self-localization}
\vspace{-10pt}
Given a specific set of lattice distortion coefficients $B_{\textbf{Q}mn}$, the effective polaron Hamiltonian can be obtained from Eq.~(\ref{H_tilde_ta}) using the thermally averaged polaron hopping and pl-ph coupling in Eq.~(\ref{polaron_parameters}). We restrict the distortion coefficients to be local, and set them to
\begin{align}
\label{ansatz}
B_{\textbf{Q}mn} =  g_{\textbf{Q}mn}\, \delta_{mn}.
\end{align}
Using this ansatz, the thermal average of the transformed matrix can be written as in Eq.~(\ref{thermal_average_M_tilde_local}), with the exponent $\lambda_{mn}(T)$ given in Eq.~(\ref{thermal_average_lambda_local}), which depends on temperature $T$ via the thermal phonon occupations $N_{\textbf{Q}}(T)$ and on the difference between the local $e$-ph coupling at the $m$ and $n$ WF sites:
\begin{gather}
\label{thermal_average_M_tilde_local_1}
\lambda_{mn} (T) = \frac{1}{N_{\Omega}}\sum_{\textbf{Q}}
        \left( N_{\textbf{Q}}(T) + \frac{1}{2} \right)
        \big| g_{\textbf{Q}mm} - g_{\textbf{Q}nn} \big|^2.
\end{gather}
%
The diagonal part of $\lambda_{mn}$ vanishes, and thus $\textrm{exp}(-\lambda_{mm})\!=\!1$ for all sites $m$. In ionic materials, usually the off-diagonal part of $\textrm{exp}(-\lambda_{mn})$ is orders of magnitude smaller than unity (typically of order $10^{-2}$ to $10^{-10}$ at 300 K), as we verify explicitly with numerical calculations here and in Ref.~\cite{Lee2021}.
In this case, polaron hopping is negligible, and we have
\begin{align}
\label{delta_identity}
\textrm{exp}(-\lambda_{mn}) \approx \delta_{mn}.
\end{align}
\indent
Substituting Eqs.~(\ref{thermal_average_M_tilde_local}), (\ref{ansatz}) and (\ref{delta_identity}) into Eq.~(\ref{polaron_parameters}), we obtain the key result for materials with negligible polaron hopping:
\begin{gather}
\label{polaron_onsite_energy}
E_{mn}=\Big( \varepsilon_{mm} - \frac{1}{N_{\Omega}}
\sum_{\textbf{Q}}\omega_{\textbf{Q}}
\big| g_{\textbf{Q}mm} \big|^2 \Big) \delta_{mn}
,\\
\label{condition_0_G}
G_{\textbf{Q}mn} = g_{\textbf{Q}mn}\delta_{mn} -B_{\textbf{Q}mn} = 0.
\end{gather}
The first equation gives the on-site polaron energy $E_{mm}$ as the sum of the electronic energy $\varepsilon_{mm}$ of the WF describing the polaron wave function and the potential energy decrease due to the lattice distortion, analogous to the CHO case [compare the second terms in Eqs.~(\ref{transformed_H_CHO}) and (\ref{polaron_onsite_energy})]. This equation further implies that the operator $a^{\dagger}_m$ in the polaron Hamiltonian in Eq.~(\ref{H_tilde_ta}) creates a self-localized polaron, as hopping to nearby sites is negligible due to the vanishing off-diagonal $E_{mn}$ hopping amplitudes.
The second equation implies that this small polaron is decoupled from all phonon modes as $G_{\textbf{Q}mn} \!=\! 0$.
Our previous work employed Eq.~(\ref{polaron_onsite_energy}) to compute the polaron energy in various families of ionic materials. In these systems, the band narrowing factor can be approximated as $e^{-\lambda_{mn}} \approx \delta_{mn} $, which implies a polaron with a very large effective mass and an ideally flat polaron band.
\\
\indent
In more weakly polar materials where polaron hopping is non-negligible, the canonical-transformation formalism is still valid, and as we show here it enables calculations of the polaron band structure. 
In this more general case, the off-diagonal elements of $e^{-\lambda_{mn}}$ cannot be neglected, and the polaron Hamiltonian matrix $E_{mn}$ can be obtained from Eq.~(\ref{polaron_parameters}), using the ansatz in Eq.~(\ref{ansatz}) and the thermal averages in Eq.~(\ref{thermal_average_M_tilde_local}). We obtain:
\begin{align}\label{general_hopping}
    E_{mn} &=\left( {\varepsilon}_{mn} -  \frac{2}{N_{\Omega}}\sum_{\textbf{Q}} \omega_{\textbf{Q}} g_{-\textbf{Q}mm}
    g_{\textbf{Q}mn} \right) e^{-\lambda_{mn}}
    \nonumber \\
     & \,\,\,\,\,\,\,\, +  \frac{1}{N_{\Omega}}\sum_{\textbf{Q}} \omega_{\textbf{Q}}  |g_{\textbf{Q}mm}|^2\delta_{mn} \,,
     \nonumber\\
\end{align}
where the on-site polaron energies $E_{mm}$ are the same as in Eq.~(\ref{polaron_onsite_energy}), but now we also compute the off-diagonal elements, namely the inter-site hopping amplitudes $E_{mn}$ (with $m \neq n$).
Setting up a tight-binding model based on this polaron Hamiltonian matrix allows us to calculate the full polaron band structure.\\

\subsection{Small polaron formation}
\vspace{-10pt}
The formation of a self-localized charge carrier in a crystal depends on two competing energies: the kinetic energy increase resulting from localizing the electronic wave function, and the energy decrease from the lattice relaxation around the charge carrier. In our formalism, this competition is clearly seen in the polaron energy in Eq.~(\ref{polaron_onsite_energy}), where $\varepsilon_{mm}$ is the electronic contribution to the polaron energy and the negative term proportional to $\sum_\mathbf{Q} \omega_\mathbf{Q} |g_{\mathbf{Q}mn}|^2$ is the energy decrease from the lattice relaxation.
\\
\indent
Computing the on-site polaron energy $E_{mm}$ allows us to predict whether a self-localized polaron will form in a material: if $E_{mm}$ is lower than the conduction band minimum (CBM) for an electron carrier, or higher than the valence band maximum (VBM) for a hole carrier, then the self-localized polaron is energetically more favorable than a delocalized Bloch state. In this scenario, the electron or hole carrier forms a small polaron and becomes self-trapped by the lattice distortion. The polaron formation energy is computed as the difference between the polaron energy $E_{mm}$ and the respective band edge; thus the formation energy for an electron polaron is
\begin{gather}
\label{our_formation_energy_1}
\Delta E_f = E_{mm} - \varepsilon_{\textrm{CBM}},
\end{gather}
and for a hole polaron
\begin{gather}
\label{our_formation_energy_2}
\Delta E_f = \varepsilon_{\textrm{VBM}} - E_{mm},
\end{gather}
where in both cases $\Delta E_f < 0$ means that polaron formation is energetically favorable.
%
%
\section{Numerical calculations}
\label{sec:numerical}
\vspace{-6pt}
\subsection{Workflow}
\vspace{-10pt}
In the canonical transformation formalism, calculations of small-polaron energies and wave functions are straightforward. In the first step, we generate maximally-localized WFs from the electronic band structure~\cite{Marzari2014}, and then calculate the $e$-ph matrix elements in the Wannier basis using the standard workflow~\cite{Perturbo}.
The next step consists in verifying numerically the approximation of negligible hopping $E_{mn} \approx E_{mm}\delta_{mn}$ (or equivalently $e^{-\lambda_{mn}} \approx \delta_{mn}$), which typically works well for strongly polar materials with self-localized polarons.
When this approximation holds, we evaluate the on-site polaron energies $E_{mm}$ using Eq.~(\ref{polaron_onsite_energy}), and then obtain the polaron formation energy using Eqs.~(\ref{our_formation_energy_1})-(\ref{our_formation_energy_2}). Note that computing $E_{mm}$ with Eq.~(\ref{polaron_onsite_energy}) is a simple post-processing of the $e$-ph calculations; it can be carried out with minimal computational cost (tens of CPU core-hours) using the {\sc {Perturbo}} code~\cite{Perturbo}.
\\
\indent
In materials where the off-diagonal elements of $E_{mn}$ and $e^{-\lambda_{mn}}$ are non-negligible, we evaluate the full effective polaron Hamiltonian matrix $E_{mn}$ using Eq.~(\ref{general_hopping}), still with minimal computational cost.
Starting from the Hamiltonian matrix $E_{mn}$, we calculate the polaron band structure using a standard tight-binding approach.
Due to its simple workflow, the canonical-transformation method enables rapid calculations of small-polaron energies in a wide range of materials~\cite{Lee2021}, and is particularly promising for high-throughput and data-driven studies of small polarons.
\begin{figure*}[!ht]
\includegraphics[width=1.01\linewidth]{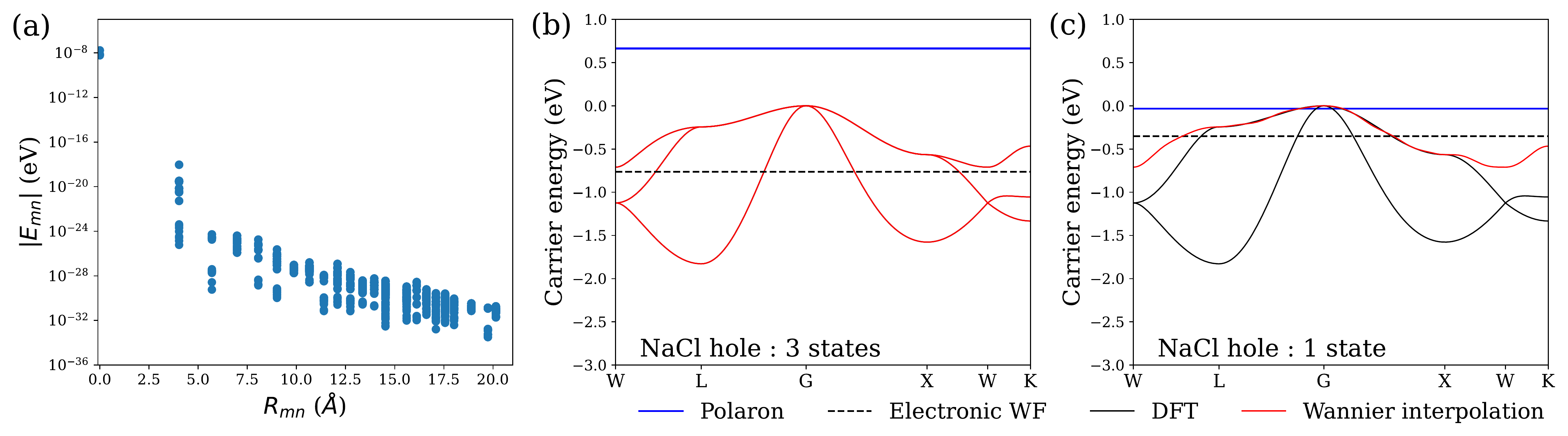}
\caption{(a) Computed polaron hopping amplitudes $E_{mn}$ for hole carriers in NaCl, shown as a function of distance $R_{mn}$ between the $m$ and $n$ WF sites at $T = 0$~K. 
(b)-(c) Calculated polaron energy for holes in NaCl, obtained by wannierizing (b) three bands and (c) one band.
Blue lines are polaron on-site energies $E_{mm}$ and dashed black lines are electronic WF energies $\varepsilon_{mm}$ [see Eq.~(\ref{polaron_onsite_energy})]. The solid black curves are the DFT band structure and the red curves the Wannier-interpolated bands, equal in number to the WFs used in the polaron calculation. The energy zero is set to the VBM. }
\label{NaCl-polaron}
\end{figure*}

\subsection{Computational details}
\vspace{-10pt}
We carry out numerical calculations on two paradigmatic systems with polarons, a simple ionic material (NaCl) and an organic crystal (naphthalene). For these case studies, we show calculations of polaron hopping amplitudes and polaron band structures, and discuss the choice of WFs to compute the polaron energy.
\\
\indent
We carry out plane-wave DFT calculations on NaCl using the {\sc Quantum ESPRESSO} code \cite{QE-2009} with norm-conserving pseudopotentials \cite{Troullier1991,Hamann2013}
and the Perdew-Burke-Ernzerhof generalized gradient approximation \cite{Perdew1996}. We use a kinetic energy cutoff of 100 Ry, an $8 \times 8 \times 8$ \textbf{k}-point grid and relaxed lattice parameters in all DFT calculations. Density functional perturbation theory \cite{Giannozzi2001} is employed to compute phonon frequencies and eigenvectors on a coarse $8 \times 8 \times 8$ \textbf{q}-point grid. The $e$-ph coupling matrix elements are first computed in the Bloch basis and then transformed to their Wannier basis counterparts, $g_{\textbf{Q}mn}$ defined above, using the {\sc Perturbo} code \cite{Perturbo} with WFs generated from {\sc Wannier90}~\cite{Marzari2014}. The calculations on napthalene follow the same workflow, using settings and numerical details provided in Ref.~\cite{Chang2022}.

\section{Results}
\subsection{Polaron hopping amplitude}
\vspace{-10pt}
The polaron energy in the canonical transformation method can be easily computed using Eq.~(\ref{polaron_onsite_energy}). Yet, to use that formula one first needs to verify that
the inter-site polaron hopping amplitude in Eq.~(\ref{general_hopping}) is negligible, so that  $E_{mn} = E_{mm}\delta_{mn}$, as a result of a diagonal band-narrowing factor in Eq.~(\ref{general_hopping}), $e^{-\lambda_{mn}} \!=\! \delta_{mn}$.
To that end, we compute $\lambda_{mn} (T)$ by carrying out the numerical integration in Eq.~(\ref{thermal_average_M_tilde_local_1}), and then obtain $E_{mn}$ using Eq.~(\ref{general_hopping}).
Recall that for the diagonal entries with $m=n$ we have $e^{-\lambda_{mm}} = 1$ by definition.
Therefore the key questions are how $E_{mn}$ decays with inter-site distance and whether its off-diagonal entries are small enough to approximate $E_{mn} = E_{mm}\delta_{mn}$ as in our recent work~\cite{Lee2021}.
\\
\indent
Figure~\ref{NaCl-polaron}(a) shows the computed polaron hopping amplitudes $E_{mn}$ for hole carriers in NaCl as a function of distance between the $m$ and $n$ WF sites. The results are given at $T \!=\! 0$~K; as the hopping amplitudes decrease monotonically with temperature, these results are an upper bound to the finite temperature hopping values.
We find that even at zero temperature the off-diagonal matrix elements of $E_{mn}$ are smaller than $10^{-6}$~eV, and they further decrease with inter-site distance and temperature.
Therefore our approach predicts that in NaCl the hole polaron is self-localized and associated with a flat polaron band.
%
%
%
%
\subsection{Choice of Wannier functions}
\vspace{-10pt}
Due to the small value of the off-diagonal hopping energies $E_{mn}$ for holes in NaCl, we can compute the hole polaron energy using Eq.~(\ref{polaron_onsite_energy}).
The first step in this calculation is the generation of WFs that accurately interpolate the band structure.
However, the choice of WFs is not unique and is a subtle point in our canonical transformation approach. As the goal is to find the lowest-energy polaron state, one could test various choices of WFs, both by changing the WF generation parameters and by wannierizing a different number of bands~\cite{Marzari2014}. Different WFs will lead to relatively small changes of polaron energy, within $\sim$1 eV based on our tests.
Therefore, if one finds an electron polaron with energy lower than the CBM (or a hole polaron with energy higher than the VBM), then our method guarantees the existence of a self-localized polaron. Conversely, if a stable self-localized polaron is not found, but the polaron energy is within $\sim$1 eV of the band edge, it's still possible that a different choice of WFs will lead to a self-localized polaron.
\\
\indent

We illustrate the role of different trial wave functions using NaCl as an example. We compare two calculations of the hole polaron energy in NaCl by generating WFs for three bands [Fig.~\ref{NaCl-polaron}(b)] or only one band [Fig.~\ref{NaCl-polaron}(c)], respectively.
The calculation using three bands gives polaron energies above the VBM and thus correctly predicts that holes in NaCl form a self-localized small polaron, in agreement with experiments~\cite{Castner1957}.
In the calculation using only one band, the polaron energy is just below the valence band edge, so a polaron is not predicted to form. However, the polaron energy is only 100 meV below the band edge, thus signaling the possible presence of a lower-energy polaron state, as confirmed in Fig.~\ref{NaCl-polaron}(b).
\\
\indent
The choice of WFs influences both the on-site electronic energy $\varepsilon_{mm}$ and the $e$-ph coupling $g_{\mathbf{Q}mm}$ as both contribute to the polaron energy $E_{mm}$ in Eq.~(\ref{polaron_onsite_energy}).
In the NaCl example, the three WFs used in Fig.~\ref{NaCl-polaron}(b) resemble the $p$ orbitals of Cl and are more spatially localized than the WF used in Fig.~\ref{NaCl-polaron}(c). As a result, they possess a greater on-site hole energy $|\varepsilon_{mm}|$ (i.e., a lower electronic energy $\varepsilon_{mm}$ in Fig.~\ref{NaCl-polaron}(b)) and a greater overlap with the phonon perturbation, resulting in a stronger on-site coupling and larger potential energy decrease in the second term of Eq.~(\ref{polaron_onsite_energy}), which leads to a more stable polaron state in Fig.~\ref{NaCl-polaron}(b). Note that the energy scale of these differences is only 0.5$-$1~eV. Therefore, whenever a self-localized polaron state is clearly unstable (say, by $>1$ eV), as we found in Ref.~\cite{Lee2021} for SrTiO$_3$, the result can be trusted without comparing different WFs.
Although one can generate and test many WFs, at present the use of a trial polaron wave function equal to a non-uniquely defined WF is a limitation of our method.
It can be overcome by formulating a generalized canonical-transformation approach, as we show below in Sec.~\ref{sec:Polaron_Energy_functional}.\\
%
\begin{figure}
\includegraphics[width=1.0\linewidth]{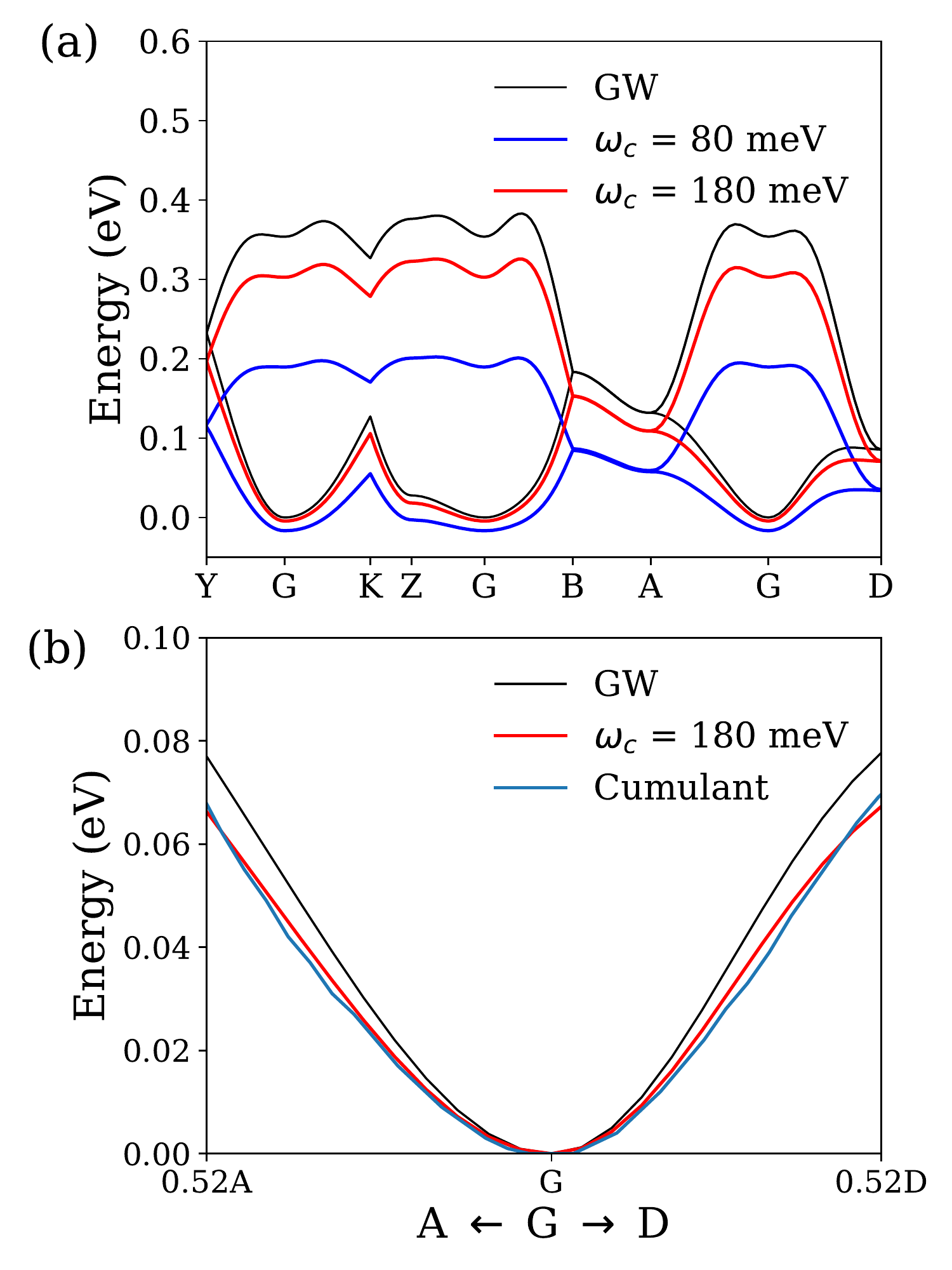}
\caption{(a) Band structure of naphthalene computed with the GW method (see Ref.~\cite{Chang2022}), shown together with the computed polaron band structure at 0~K for two different phonon cutoff energies, $\omega_c = $ 80 meV and 180 meV.
(b) Polaron band structure from the canonical transformation (with $\omega_c = $ 180 meV) compared with polaron calculations using the cumulant method (Ref.~\cite{Chang2022}).
The energy zero is the CBM.
}
\label{Naph-polaron}
\end{figure}
\indent

\subsection{Polaron band structure}
We carry out a polaron band structure calculation on naphthalene, an organic semiconductor with non-negligible polaron hopping. Electron carriers in naphthalene possess a narrow band width of $\sim$200 meV, leading to pronounced polaron effects~\cite{Chang2022}.
We wannierize the two lowest conduction bands, and calculate the polaron band structure using tight-binding with the energies $E_{mn}$ in Eq.~(\ref{general_hopping}).
In these calculations, we exclude phonons with energy lower than a cutoff $\omega_c$ when computing $\lambda_{mn}$ in Eq.~(\ref{thermal_average_M_tilde_local_1}) and  $E_{mn}$.
Excluding phonons with up to several times the electronic hopping energy $|t|$ is physically justified in the canonical transformation when hopping is present~\cite{Berkelbach2020}. As the charge carrier hops and the lattice rearranges, only phonons with frequency greater than the hopping energy can follow the charge carrier and make up its phonon cloud. Lower energy phonons contribute to the residual pl-ph interaction and can be treated as dynamical disorder~\cite{Berkelbach2020}. Here, from WF analysis we obtain a maximum electron hopping energy $|t|\approx 40$~meV.
\\ 
\indent
Figure~\ref{Naph-polaron}(a) shows the polaron band structure at 0~K for two different cutoffs, $\omega_c=$ 80 and 180 meV; the first cutoff is twice the hopping energy and the second is as high as possible but below a strongly coupled optical phonon.
For both cutoffs, the minima of the polaron bands are below the conduction band, signaling the presence of a stable polaron. Even at 0~K, the polaron bands are narrower than the electronic GW bands due to the zero-point term in $\lambda_{mn}$.
The band narrowing decreases for greater cutoff values, leaving an adjustable parameter in the theory. Guided by Holstein's work~\cite{Holstein1959}, we argue that the optimal cutoff is the highest phonon energy that permits the inclusion of strongly coupled optical phonons (here this value is 180 meV).
In Fig.~\ref{Naph-polaron}(b) we compare the low-energy polaron dispersion in the canonical transformation with $\omega_c =$ 180~meV to our recent cumulant calculation in naphthalene~\cite{Chang2022}. For this optimal cutoff, the two methods give polaron band structures in quantitative agreement $-$ both methods predict a polaron with dispersive bands and modest mass renormalization.

\clearpage
\newpage
%
%
\section{Method comparison}
\subsection{Energy functional method}
\label{sec:their_formalism}
\vspace{-10pt} 
We briefly summarize the formalism of Sio et al.~\cite{Sio201901, Sio201902} for first-principles polaron calculations. Their approach models the polaron as a \textit{single} excess charge carrier, and calculates its electronic wave function $\psi$ and associated lattice distortion $u_{\kappa}^0$ by minimizing the polaron energy functional \{see Eq.~(23) in Ref.~\cite{Sio201902}\}:
\begin{align}
\label{their_polaron_energy_functional}
E_p \left[ \psi, u^0_\kappa \right] &=
\frac{1}{2} \sum_{\kappa \kappa^{\prime}} \Phi_{\kappa \kappa^{\prime}} u^0_{\kappa} u^0_{\kappa^{\prime}} \\
&+\int d\textbf{r}\, \psi^{*}(\textbf{r}) \left( H_{\textrm{KS}} + \sum_{\kappa}
\frac{\partial V_{\textrm{KS}}}{\partial u^0_{\kappa}} u^0_{\kappa} \right) \psi (\textbf{r}), \nonumber
\end{align}
where $\kappa$ is a composite index for atoms and Cartesian coordinates (similar to Eq.~(\ref{utilde}), $\kappa\!=\!cs\alpha$), and $\Phi_{\kappa \kappa^{\prime}}$ are interatomic force constants; $H_{\textrm{KS}}$ and $V_{\textrm{KS}}$ are, respectively, the Kohn-Sham (KS) Hamiltonian and KS potential at equilibrium without the excess charge carrier.
\\
\indent
In Refs.~\cite{Sio201901, Sio201902}, the wave function $\psi$ of the excess electron is written as a superposition of Bloch states (with band index $i$ and crystal momentum $\mathbf{k}$) or WFs (with composite index $m=j_{m}\textbf{R}_m$, as above):
\begin{align}
\label{their_wf_expansion}
\psi(\textbf{r}) = \frac{1}{\sqrt{N_{\Omega}}} \sum_{i \textbf{k}} A_{i \textbf{k}} \psi_{i \textbf{k}} (\textbf{r})
= \sum_{m}A_m w_m (\textbf{r}),
\end{align}
where $A_{i\mathbf{k}}$ and $A_m$ are expansion coefficients for the wave function in the Bloch and Wannier basis, respectively.
The lattice distortion $u^0_{\kappa}$ due to the excess charge carrier is expanded in the basis of phonon eigenvectors as 
\begin{align}
\label{their_lattice_distortion}
u^0_{cs\alpha} = - \frac{2}{N_{\Omega}} \sum_{\textbf{Q}} B^{*}_{\textbf{Q}}
        \frac{1}{\sqrt{2 M_s \omega_{\textbf{Q}}}}
        e^{s\alpha}_{\textbf{Q}}e^{i\textbf{q}\cdot \textbf{R}_{c}},
\end{align}
where $B_{\mathbf{Q}}$ are scalar lattice-distortion coefficients that are independent of electronic band or WF site.
Minimizing the polaron energy functional with respect to $\psi^{*}$ and $u^0_\kappa$ gives
a set of coupled polaron equations for the coefficients $A_{i\textbf{k}}$ and $B_{\textbf{Q}}$ \{see Eqs.~(37)-(38) in Ref.~\cite{Sio201902}\}~\footnote{Note that similar to above we factored out the phonon energy and defined the $e$-ph matrix elements as
$g_{ji\nu}(\textbf{k},\textbf{q}) = \tilde{g}_{ji\nu}(\textbf{k},\textbf{q})/ (\hbar \omega_{\mathbf{Q}})$,
where $\tilde{g}_{ji\nu}(\textbf{k},\textbf{q})$ are the usual $e$-ph matrix elements in the Bloch basis~\cite{Perturbo}.}: 
\begin{gather}
\label{first_polaron_eq_k_space}
\frac{2}{N_\Omega}\sum_{i^{\prime} \mathbf{Q}} \omega_{ \mathbf{Q}} B_{ \mathbf{Q}}\,g^{*}_{i^{\prime} i \nu}(\textbf{k},\textbf{q})A_{i^{\prime} \textbf{k}+\textbf{q}}
    = (\varepsilon_{i\textbf{k}}-\varepsilon)A_{i\textbf{k}}\,\\
\label{second_polaron_eq_k_space}
B_{\mathbf{Q}}=\frac{1}{N_{k}} \sum_{i^{\prime} i \textbf{k}} A^{*}_{i^{\prime} \textbf{k}+\textbf{q}}\,
    g_{i^{\prime}i\nu}(\textbf{k},\textbf{q})
    \,A_{i\textbf{k}}
\end{gather}
which are solved self-consistently
by first assuming a set of coefficients $B_{\mathbf{Q}}$ and then solving for $A_{i\mathbf{k}}$ in the first equation.
The process is then repeated until convergence.
The resulting polaron formation energy is~\cite{Sio201902}
\begin{align}
\label{their_formation_energy}
\Delta E_f &= \varepsilon - \varepsilon_{\textrm{CBM}}+\frac{1}{2}\sum_{\kappa\kappa^{\prime}}
\Phi_{\kappa\kappa^{\prime}} u^0_{\kappa} u^0_{\kappa^{\prime}}
\nonumber
\\
&= \frac{1}{N_{\textbf{k}}}\sum_{i\textbf{k}} |A_{i\textbf{k}}|^2 (\varepsilon_{i\textbf{k}} - \varepsilon_{\textrm{CBM}}) - \frac{1}{N_\Omega}\sum_{\textbf{Q}} \omega_{\textbf{Q}}|B_{\textbf{Q}}|^2\,.
\end{align}

\subsection{Comparison I: Polaron energy}
\vspace{-10pt}

We now prove the equivalence of the small-polaron energy in the canonical transformation and energy functional methods: for a given small polaron wave function, the on-site polaron energy computed using Eq.~(\ref{polaron_onsite_energy}) in our method is identical to the polaron energy in Refs.~\cite{Sio201901,Sio201902}.
\\
\indent
In our canonical transformation formalism~\cite{Lee2021},
we use a single WF as the trial small-polaron wave function.
Suppose this WF is centered at site $0$, then the polaron wave function is
\begin{equation}
  \ket{w_0} = a^{\dagger}_0 \ket{0}.
\end{equation}
This wave function is defined in the canonical transformed Hamiltonian.
For this state, we can set all the distortion coefficients to zero except $B_{\textbf{Q}00}$, as can be seen from the expectation value of Eq.~(\ref{b_tilde}): $\langle0|a_{0} \sum_{mn} B_{\textbf{Q}mn} a^{\dagger}_m a_n a^{\dagger}_{0}|0\rangle = B_{\textbf{Q}00}$.
From Eqs.~(\ref{ansatz}) and (\ref{polaron_onsite_energy}), the on-site energy of this polaron state is
\begin{equation}
  E_{00} = {\varepsilon}_{00}  -
     \frac{1}{N_{\Omega}}\sum_{\textbf{Q}} \omega_{\textbf{Q}} |g_{\textbf{Q}00}|^2,
\end{equation}
where ${\varepsilon}_{00}$ is the WF energy and $g_{\textbf{Q}00}$ the $e$-ph matrix element in Wannier basis at the polaron site.
Both of these quantities depend on the WF choice, as discussed above.
Without loss of generality, we assume that the WF with lowest on-site polaron energy $E_{00}$ can be expanded in Bloch basis as
\begin{equation}
\ket{w_0} = \frac{1}{\sqrt{N_k}}\sum_{i\mathbf{k}}\,A_{i\mathbf{k}}\ket{\psi_{i\mathbf{k}}}.
\end{equation}
Using the transformation between Wannier and Bloch basis [see
Eq.~(\ref{Bloch2Wannier}) in  Appendix~\ref{sec:appendix:elastic-fix}],
the polaron formation energy $\Delta E_f = E_{00} - \varepsilon_{\textrm{CBM}}$ is
\begin{align}
  \Delta E_f = \frac{1}{N_k}\sum_{i\mathbf{k}} |A_{i\mathbf{k}}|^2({\varepsilon}_{i\mathbf{k}} - \varepsilon_{\textrm{CBM}}) - \frac{1}{N_{\Omega}}\sum_{\textbf{Q}} \omega_{\textbf{Q}} |g_{\mathbf{Q}00}|^2,
\end{align}
where
\begin{equation}
\label{equivalence-gQ00}
g_{\textbf{Q}00}=\frac{1}{N_k}\sum_{i'i\textbf{k}}
A^*_{i'\textbf{k}+\textbf{q}}g_{i'i\nu}(\textbf{k},\textbf{q})
A_{i\textbf{k}}.
\end{equation}

This polaron formation energy $\Delta E_{f}$ and on-site e-ph coupling $g_{\mathbf{Q}00}$ are exactly the same as, respectively, the formation energy and distortion coefficient $B_{\mathbf{Q}}$ in Refs.~\cite{Sio201901,Sio201902}.
%
More precisely, the polaron energy $E_{00}$ is identical to
the eigenvalue $\varepsilon$ of Eq.~(\ref{first_polaron_eq_k_space}),
the first polaron equation in Sio et al.~\cite{Sio201902}, provided that we add to their eigenvalue the elastic energy associated with the polaron lattice distortion, $\sum_\mathbf{Q} \omega_\mathbf{Q} |B_\mathbf{Q}|^2 = 1/2\, \sum_{\kappa \kappa'}\Phi_{\kappa \kappa'} u^0_\kappa u^0_{\kappa'}$ (see Appendix~\ref{sec:appendix:elastic-fix}).
This constant term is included in the canonical transformation polaron energy $E_{00}$, whereas in the energy functional method of Ref.~\cite{Sio201902} it is added to the eigenvalue after the calculation [see Eq.~(\ref{their_formation_energy})].
%
%
\\
\indent
Available numerical results confirm this equivalence for cases where the polaron is self-localized and well described by a single WF. For example, we recently computed the polaron formation energy for electrons in $\text{Li}_2\text{O}_2$, and obtained a value of -4.905 eV~\cite{Lee2021} that is nearly identical to the -4.87 eV value found in Ref.~\cite{Sio201902}. Note that in our canonical transformation method the polaron energy is computed straightforwardly using Eq.~(\ref{polaron_onsite_energy}), with negligible computational cost even for large systems. By contrast, the energy functional method in Refs.~\cite{Sio201901,Sio201902} requires solving an eigenvalue problem self-consistently and extrapolating the result to an infinite $\mathbf{k}$-point grid size, with significant computational cost. 
The key advantage of our approach is the use of WFs as a more natural, localized basis set to describe small polarons, which enables bypassing costly calculations in momentum space.
\\
\indent
Comparing the two methods for a more delocalized polaron wave function is more challenging.
On one hand, our approach can be generalized to take into account an arbitrary polaron wave function, giving a general canonical transformation formalism that can treat both small and large polarons (see below).
On the other hand, the method of Refs.~\cite{Sio201901,Sio201902} can already describe an arbitrary polaron wave function (as a superposition of WFs at multiple sites), but as we show below it has important limitations for addressing key polaron physics such as thermal effects and polaron localization.

\subsection{Comparison II: Polaron wave function}
\label{sec:Polaron_Energy_functional}
\vspace{-10pt}
\noindent
\textit{Canonical transformation method.} In the canonical transformation formalism, it is not obvious how to determine the polaron wave function~\cite{Lee2021}. While Eq.~(\ref{polaron_onsite_energy}) gives the polaron energy for a WF localized at site $m$, the choice of this WF is not unique.
Different WFs may lead to different electronic energies $\varepsilon_{mm}$ and $e$-ph interactions $g_{\mathbf{Q}mm}$, and thus different polaron energies $E_{mm}$.
The most stable polaron state corresponds to the wave function minimizing the polaron energy, but that wave function may be a nontrivial combination of WFs.
\\ 
\indent
%
%
We present an extension to our canonical transformation formalism using a general polaron wave function written as a superposition of WFs,
\begin{equation}
  \ket{\psi} = \sum_{m}A_{m}a^\dagger_{m}\ket{0} = \sum_m A_m \ket{w_m}
\end{equation}
with normalization $ \sum_m |A_m|^2 = 1$.
The energy of this polaron state in this generalized canonical-transformation formalism is
  \begin{align}
  \nonumber
  E[A,B] &= {\braket{\psi|\tilde{H}|\psi}}= \sum_{mn}A^*_m A_n\braket{0|a^\dagger_m\tilde{H}a_n|0}
  \\
   \label{our_energy_functional}
  &=\frac{1}{N_{\Omega}} \sum_m |A_m|^2 \sum_{\mathbf{Q}} \omega_{\mathbf{Q}} |B_{\mathbf{Q}mm}|^2
  \\
   \nonumber
  &  \!\!\!\!\!\!\!\!\!\!+  \sum_{mn}A_m^*A_n \,e^{-\lambda_{mn}}\!\! \left( \epsilon_{mn} -\frac{2}{N_\Omega}\sum_{\mathbf{Q}}\omega_\mathbf{Q} B_{-\mathbf{Q}mm} \, g_{\mathbf{Q}mn}  \right)\!,
  \end{align}
where we ignored the off-diagonal matrix elements of $B_{\mathbf{Q}mn}$.
This energy functional depends on the Wannier-basis coefficients $A_m$ of the electronic wave function
and on the lattice distortion coefficients $B_{\mathbf{Q}mm}$ describing how the lattice responds to the charge carrier.
\\
\indent
The resulting full wave function $\ket{\Phi}$ of the combined electron-plus-phonon system has a general form with \mbox{\textit{entangled}} electrons and phonons.
In the canonical transformation framework, the full wave function in the original Hamiltonian is
\begin{align}
    \label{our_variational_wf}
   \ket{\Phi} &= e^{S}\ket{\psi} \\
   \nonumber
   &= \sum_{m} A_m\, e^{\frac{1}{\sqrt{N_{\Omega}}}\sum_{\mathbf{Q}}B_{\mathbf{Q}mm}(b^{\dagger}_\mathbf{Q} - b_{-\mathbf{Q}})} a^{\dagger}_m \ket{0}
\end{align}
where $S$ is defined in Eq.~(\ref{polaron_S}) and we neglected the off-diagonal matrix elements of $B_{\mathbf{Q}mn}$.
\\
\indent
Full minimization of the energy functional in Eq.~(\ref{our_energy_functional}) with respect to $A_m$ and $B_{\mathbf{Q}mm}$ is challenging and will be pursued in future work.
For materials with strong $e$-ph interactions leading to self-localized small polarons, the ansatz introduced above of a polaron localized at a single site that is free of hopping (obtained respectively by setting $A_n = \delta_{n0}$ and $B_{\mathbf{Q}mn} \!=\! g_{\mathbf{Q}mm} \delta_{mn}$) is a good approximation to the stationary solution of the generalized canonical-transformation functional in Eq.~(\ref{our_energy_functional}).
%
In that approximation, the polaron wave function reduces to the tensor product state
\begin{equation}
 \label{tps-our}
 \ket{\Phi} = e^{\frac{1}{\sqrt{N_{\Omega}}}\sum_{\mathbf{Q}}g_{\mathbf{Q}00}(b^{\dagger}_\mathbf{Q} - b_{-\mathbf{Q}})}
 a^{\dagger}_0\ket{0}
\end{equation}
defined uniquely by one distortion coefficient for each phonon mode, $B_{\mathbf{Q}} = g_{\mathbf{Q}00}$.
For a more general polaron state, if one uses site-dependent distortion coefficients $B_{\mathbf{Q}mm}$ together with an electronic wave function with amplitude at multiple sites, then the full wave function $\ket{\Phi}$ entangles the electrons and phonons, and includes both the solution in Eq.~(\ref{tps-our}) and the approach by Sio et al.~\cite{Sio201902} as subcases, as we show below.
\\
%
%
\textit{Energy functional method.}
In the formalism of Refs.~\cite{Sio201901, Sio201902}, the full wave function is effectively a tensor product of the electron and phonon wave functions:
\begin{align}
  \label{their_variational_wf}
  \ket{\Phi} &= \left(\sum_{m}A_m a^{\dagger}_m \ket{0}_{e} \right)\otimes
  \left(e^{\frac{1}{\sqrt{N_{\Omega}} }\sum_{\mathbf{Q}}B_\mathbf{Q}(b^{\dagger}_\mathbf{Q} -b_{-\mathbf{Q}})} \ket{0}_p \right)
  \\
  \nonumber
  &= \sum_{m} A_m e^{\frac{1}{\sqrt{N_{\Omega}} }\sum_{\mathbf{Q}}B_\mathbf{Q}(b^{\dagger}_\mathbf{Q} - b_{-\mathbf{Q}})}a^{\dagger}_m \ket{0},
\end{align}
where $\ket{0}_e$ and $\ket{0}_p$ are the electron and phonon vacuums, respectively, and $\ket{0}\!=\! \ket{0}_p\!\ket{0}_e$ the combined vacuum.
\\
\indent
We briefly outline the derivation of this result. The phonon state in Ref.~\cite{Sio201901,Sio201902} is characterized only by the classical displacements $u_{\kappa}^0$.
As discussed at the end of in Sec.~\ref{sec:toy_model}, the ground state of a shifted harmonic oscillator centered at $u^0 = \frac{B}{\sqrt{2 m \omega}}$ can be described by the coherent state~\cite{SHI2018245} %
\begin{equation}
\ket{u^0} = e^{-\frac{B}{2}(b^\dagger - b)}\ket{0}_{p}.
\end{equation}
We can extend this state to describe the entire distorted lattice, with the position of each atom $\kappa$ displaced by $u_\kappa^0$ as in Eq.~(\ref{their_lattice_distortion}):
\begin{equation}
   \ket{\{u^0_\kappa\}} =  e^{\frac{1}{\sqrt{N_{\Omega}} }\sum_{\mathbf{Q}}B_\mathbf{Q}[A](b^{\dagger}_\mathbf{Q}-b_{-\mathbf{Q}})} \ket{0}_{p},
\end{equation} 
from which one obtains the full wave function in Eq.~(\ref{their_variational_wf}) for the approach in Refs.~\cite{Sio201901,Sio201902}. Here, we noted explicitly that the distortion coefficients $B_\mathbf{Q}[A]$ are functionals of the wave function coefficients $A_m$ via the second polaron equation in Wannier form [see Eq.~(\ref{second_polaron_eq})].
\\
\indent
Therefore, while the electrons and phonons are disentangled as a tensor-product state in the energy-functional method, they are still coupled in a mean-field sense via the distortion coefficients $B_\mathbf{Q}[A]$ from the second polaron equation, which describes the mean-field effect of the electrons and $e$-ph coupling on the lattice.
Starting from our wave function in Eq.~(\ref{our_variational_wf}), if we make the canonical-transformation distortion coefficients $B_{\mathbf{Q}mm}$ independent of site and electron orbital index $m$, $B_{\mathbf{Q}mm} \rightarrow B_{\mathbf{Q}}$,
we obtain the variational ansatz in Eq.~(\ref{their_variational_wf}) as a special case of the canonical transformation formalism.
%
Similarly, the polaron energy functional in Refs.~\cite{Sio201901,Sio201902} can be viewed a special case of the generalized canonical transformation approach given in Eq.~(\ref{our_energy_functional}).
\\
\indent
%
%
\subsection{Comparison III: Temperature dependence}
\vspace{-10pt}
Thermal effects are essential in polaron physics.
For small-polarons, the polaron bands become progressively flatter as the temperature increases, until the charge carrier ultimately self-localizes. In the canonical transformation formalism, these effects are encoded in the temperature dependent band-narrowing factor, $\exp[-\lambda_{mn}(T)]$.
Analysis of the exponent $\lambda_{mn}(T)$ in Eq.~(\ref{thermal_average_lambda_local}) shows that the band-narrowing factor decreases with temperature due to an increase in the phonon occupations $N_{\mathbf{Q}}$, leading to a progressive flattening of the polaron bands, consistent with the picture discussed above. Even at $T\!=\!0$~K, where the phonon occupations $N_{\mathbf{Q}}$ vanish, the exponent $\lambda_{mn}(T)$ can still be relatively large due to the zero-point motion of the lattice, resulting in a significant zero temperature band-narrowing factor:
\begin{equation}
    \label{zpm}
    e^{-\lambda_{mn} } (T=0) = e^{-\frac{1}{2 N_\Omega} \sum_{\mathbf{Q}} \left| B_{\mathbf{Q}mm} - B_{\mathbf{Q}nn} \right|^2}.
\end{equation}
%
This result shows that there is a finite zero-point
polaron band renormalization due to the site and WF dependence of the distortion coefficients. 

\indent
In the energy functional approach of Refs.~\cite{Sio201901,Sio201902}, these thermal and zero-point effects are missing entirely, an important limitation for a polaron theory.
This point is clearly seen in their energy functional in Eq.~(\ref{their_polaron_energy_functional}), which includes only the static lattice distortion $u^0_\kappa$ but no terms associated with lattice vibrations. It can be better understood by comparing their first polaron equation in Wannier form, Eq.~(\ref{first_polaron_eq}), with the corresponding term in the generalized canonical-transformation functional, the third line in Eq.~(\ref{our_energy_functional}).
This comparison shows that the approach by Sio et al.~\cite{Sio201901,Sio201902} is equivalent to assuming $e^{\lambda_{mn}} =1$, or equivalently $\lambda_{mn} = 0$, which neglects both the zero- and finite-temperature polaron band narrowing.
%
Their lack of band narrowing and temperature dependence
is a consequence of not including the lattice vibrations, which is equivalent to setting $N_{\mathbf{Q}}+1/2 = 0$ in Eq.~(\ref{thermal_average_M_tilde_local_1}) for the canonical transformation formalism.
Note also that the energy functional method in Refs.~\cite{Sio201901,Sio201902} describes the polaron as an isolated system (essentially, a localized defect), and thus a polaron band structure is missing altogether. This is why that method cannot be extended straightforwardly to include thermal effects on the polaron band structure and effective mass.
\\
\indent
%
%
The effect of temperature is also critical for polaron dynamics. A polaron can hop from between different sites assisted by the thermal lattice vibrations, and the distortion gets transferred to the new site~\cite{Emin1982}. In many materials with polaron effects, as temperature increases charge transport transitions from a band-like mechanism to thermally-activated charge hopping. Due to its thermally activated nature, describing charge hopping requires distortion coefficients that depend on site and electronic state.
In the canonical transformation formalism, the distortion coefficients $B_{\mathbf{Q}mn}$ are associated with electronic hopping amplitudes between WF sites $m$ and $n$, via terms proportional to $B_{\mathbf{Q}mn}a^\dagger_m a_n$ that couple explicitly the electron and lattice dynamics.
By contrast, the energy functional formalism, using site- and electronic state-independent distortion coefficients $B_{\mathbf{Q}}$~\cite{Sio201901,Sio201902}, couples the electron and lattice dynamics in a mean field way, as seen in the second polaron equation in Wannier basis, Eq.~(\ref{second_polaron_eq}).
\\
\indent

\subsection{Comparison IV: Polaron localization}
\vspace{-10pt}
In the canonical transformation method, polaron self-localization is easy to verify starting from the effective Hamiltonian. When the condition $e^{-\lambda_{mn}(T)} \approx \delta_{mn}$ is satisfied, the Hamiltonian reduces to the diagonal matrix $E_{mn}$ in Eq.~(\ref{polaron_onsite_energy}), and thus the polaron is localized at a single site with a vanishing hopping amplitude.
In this scenario, when the canonical transformation method predicts a polaron on-site energy lower than the band edge, we conclude that the formation of a self-localized polaron with a nearly flat polaron band is energetically favorable.
In materials with non-negligible polaron hopping, one can use Eq.~(\ref{general_hopping}) in the canonical transformation approach to compute the temperature dependent polaron band structure.
\\
\indent
To guarantee that these polaron band structure calculations are physically meaningful, the polaron Hamiltonian matrix $E_{mn}$ in Eq.~(\ref{general_hopping}) needs to have the same translation symmetry as the lattice.
This translational invariance is simple to show in our canonical transformation formalism.
Recall that the WF index $n$ is a composite index labeling both the site and WF, $n=j_n\textbf{R}_n$.
We translate the $m$ and $n$ WFs by a lattice vector $\mathbf{R}$, shifting them to new sites $m'=j_m \textbf{R}_m+\mathbf{R} $ and $n'=j_n \textbf{R}_n+\mathbf{R} $.
Using Eq.~(\ref{Bloch2Wannier}), the translated $e$-ph coupling matrix elements become
$ g_{\textbf{Q}m'n'} = g_{\textbf{Q}mn}\,e^{i\textbf{q}\cdot \mathbf{R}}$,
and using this relation to evaluate the translated Hamiltonian matrix $E_{m'n'}$ in Eq.~(\ref{general_hopping}),
one obtains the translational invariance condition $E_{m'n'} = E_{mn}$.
\\
\indent
The situation is different in the energy functional method~\cite{Sio201901,Sio201902}, where the polaron is described as an isolated system consisting of a single charge carrier plus a lattice distortion around it.
The distorted lattice induces a local potential which explicitly breaks the translational symmetry. As a result, a polaron band structure cannot be defined at any temperature, and polaron self-localization is not deduced from a vanishing polaron hopping or bandwidth.
Rather, in the energy functional method, polaron self-localization is inferred from the presence of a bound state in a static potential generated by the distorted lattice~\cite{Sio201901,Sio201902}.
This approach hides the complex physics of the polaron problem, with key temperature dependent inter-site hopping, and treats it as a simple quantum mechanical problem of a particle in a localized potential. Yet, in reality the attractive potential felt by the excess electron is neither static nor temperature independent, as it is determined by the zero-point and thermal motion of the lattice.
\\
\indent
For small polarons, self-localization cannot be guaranteed by the presence of a bound state in the method of Refs.~\cite{Sio201901,Sio201902}, unless one can prove a negligible inter-site hopping and extend the method to finite temperatures.
%
For materials with non-negligible polaron hopping, or where polaron localization varies significantly with temperature, the lack of translation symmetry, hopping amplitude, and thermal effects in the energy functional method of Refs.~\cite{Sio201901,Sio201902} currently prevents quantitative comparisons with our canonical transformation approach.\\

\section{Conclusion}
\vspace{-10pt}
\indent
This work analyzes and compares two methods that advance first-principles studies of polarons. Both methods can compute the polaron energy and lattice distortion with calculations that use only a unit cell of the material. These approaches leverage \textit{ab initio} $e$-ph calculations and related software packages to carry them out efficiently on modern computer architectures.  
We have highlighted the proper treatment of thermal effects, translational invariance, and polaron self-localization in the canonical transformation framework.
\\
\indent
We believe that more work is needed to bring our canonical transformation method to full fruition. We have shown that it can be extended to explicitly compute the polaron wave function and treat both small and more delocalized polarons. As it includes an explicit coupling of charge hopping and lattice distortion, the canonical transformation method can also be extended, using linear-response theory, to study charge transport in the polaron hopping regime. 
Our analysis highlighted a common root for the canonical transformation and energy functional methods, suggesting that proper extensions of both approaches will enable exciting future developments in polaron physics.

\section*{Acknowledgements}
\vspace{-10pt}
The authors thank Nien-En Lee for fruitful discussions. This work was supported by the Air Force Office of Scientific Research through the Young Investigator Program, Grant FA9550-18-1-0280. M.B. was partially supported by the Liquid Sunlight Alliance, which is supported by the U.S. Department of Energy, Office of Science, Office of Basic Energy Sciences, under Award No. DE-SC0021266.
This research used resources of the National Energy Research Scientific Computing Center (NERSC), a U.S. Department of Energy Office of Science User Facility located at Lawrence Berkeley National Laboratory, operated under Contract No. DE-AC02-05CH11231.

\appendix
\section{}
\vspace{-10pt}
\label{sec:appendix:elastic-fix}
To compare the polaron energy in the canonical transformation and energy functional methods, we use the identity~\cite{Marzari2014}
%
\begin{align}
A_{i\textbf{k}}=\sum_{j_m \textbf{R}_m} e^{-i \textbf{k} \cdot \textbf{R}_m} \, \mathcal{U}^{\textbf{k}}_{ij_m} \, A_{j_m\textbf{R}_m}\nonumber
\end{align}
to replace $A_{i\textbf{k}}$ in Eqs.~(\ref{first_polaron_eq_k_space}) and (\ref{second_polaron_eq_k_space}), where $\mathcal{U}^{\textbf{k}}_{ij_m}$ is the unitary matrix from the WF generation process~\cite{Marzari2014}.
The two polaron equations of Ref.~\cite{Sio201902}, rewritten in the Wannier basis, become
\begin{gather}
\label{first_polaron_eq}
\sum_{mn}A^{*}_m A_n \Big[
\varepsilon_{mn} - \frac{2}{N_{\Omega}}\sum_{\textbf{Q}}\omega_{\textbf{Q}}\, B_{-\textbf{Q}}\, g_{\textbf{Q}mn}
\Big] = \varepsilon, \\[7pt]
\label{second_polaron_eq}
B_{\textbf{Q}}=\sum_{mn} A^{*}_m g_{\textbf{Q}mn} A_n,
\end{gather}
where we have used the following definitions for the real-space electron hopping amplitude and $e$-ph coupling constants:
\begin{gather}
\varepsilon_{mn} = \frac{1}{N_\Omega} \sum_{i\textbf{k}} e^{-i \textbf{k} \cdot (\textbf{R}_n-\textbf{R}_m)}
\mathcal{U}^{\dagger\textbf{k}}_{j_m i } \, \varepsilon_{i\textbf{k}} \, \mathcal{U}^{\textbf{k}}_{i j_n},\label{Bloch2Wannier}
\\
g_{\textbf{Q}mn}=\frac{1}{N_\Omega}\sum_{i^{\prime}i\textbf{k}} e^{i(\textbf{k}+\textbf{q})\cdot \textbf{R}_m}
\mathcal{U}^{\dagger\textbf{k}+\textbf{q}}_{i^{\prime}j_m }g_{i^{\prime}i\nu}(\textbf{k},\textbf{q})
\mathcal{U}^{\textbf{k}}_{ij_n}e^{-i\textbf{k}\cdot \textbf{R}_{n}}. \nonumber
\end{gather}
\\
\indent
The equality used above for the elastic energy,
\begin{align}
\frac{1}{2} \sum_{\kappa \kappa^{\prime}} \Phi_{\kappa \kappa^{\prime}} u^0_{\kappa} u^0_{\kappa^{\prime}}
=\frac{1}{N_{\Omega}}\sum_{\textbf{Q}}\omega_{\textbf{Q}}
        |B_{\textbf{Q}}|^2,
\end{align}
can be derived by substituting in the left-hand side the expression for $u^0_{\kappa}$ in Eq.~(\ref{their_lattice_distortion}) and using the identities:
\begin{gather}
\Phi_{\kappa \kappa^{\prime}} = \frac{\sqrt{M_s M_{s^{\prime}}}}{N_{\Omega}}
\sum_{\textbf{q}}e^{-i \textbf{q} \cdot(\textbf{R}_{c^\prime}-\textbf{R}_c)}D_{s\alpha\textrm{,}s^{\prime}\alpha^{\prime}}(\textbf{q}),
\nonumber\\
\sum_{s^{\prime}\alpha^{\prime}}D_{s\alpha\textrm{,}s^{\prime}\alpha^{\prime}}(\textbf{q})
e^{s^{\prime}\alpha^{\prime}}_{\textbf{Q}}=\omega^2_{\textbf{Q}} e^{s\alpha}_{\textbf{Q}},
\nonumber\\
\sum_{s\alpha} e^{* s\alpha}_{\nu\textbf{q}}e^{s\alpha}_{\nu^{\prime}\textbf{q}} = \delta_{\nu\nu^{\prime}}
\nonumber
\end{gather}
where $D_{s\alpha\textrm{,}s^{\prime}\alpha^{\prime}}(\textbf{q})$ is the dynamical matrix.
\\
\indent
A subtle question is why the polaron energy includes this elastic energy term in the canonical transformation but not in the the energy functional approach, where it needs to be added to the eigenvalue of the first polaron equation [see Eq.~(\ref{their_formation_energy})].
A correction to the energy functional in Refs.~\cite{Sio201901,Sio201902} allows us to properly include the elastic energy in the polaron equation eigenvalue.
The functional in Eq.~(\ref{their_polaron_energy_functional}) suffers from an inaccuracy: when the excess charge vanishes, as can be obtained by setting the polaron wave function $\psi = 0$, the polaron energy does not vanish: incorrectly, it equals the elastic energy induced by the polaron.
This unphysical behavior can be addressed by properly coupling the elastic energy with its source, the charge distribution $|\psi(\mathbf{r})|^2$, in the energy functional, changing the first term in Eq.~(\ref{their_polaron_energy_functional}) as
\begin{align}
\frac{1}{2} \sum_{\kappa\kappa^{\prime}}\Phi_{\kappa\kappa^{\prime}} u^0_{\kappa}u^0_{\kappa^{\prime}}
\rightarrow
\frac{1}{2} \sum_{\kappa\kappa^{\prime}}\Phi_{\kappa\kappa^{\prime}} u^0_{\kappa}u^0_{\kappa^{\prime}}
\int d\textbf{r} \big| \psi(\textbf{r}) \big|^2.
\end{align}
This way, $\psi = 0$ gives a polaron energy $E_p \!=\! 0$, and the revised energy functional becomes:
\begin{align}
\label{our_polaron_energy_functional}
&E_p \left[ \psi, u^0_\kappa \right] =  \int d\textbf{r}\, \psi^{*}(\textbf{r}) \times
\\
&\qquad\quad\left( \frac{1}{2} \sum_{\kappa \kappa^{\prime}} \Phi_{\kappa \kappa^{\prime}} u^0_{\kappa} u^0_{\kappa^{\prime}}
+ \sum_{\kappa} \frac{\partial V_{\textrm{KS}}}{\partial u^0_{\kappa}} u^0_{\kappa} + H_{\textrm{KS}} \right) \psi (\textbf{r}). \nonumber
\end{align}

Varying with respect to the polaron wave function, under the constraint of its normalization, gives a revised first polaron equation that properly includes the elastic energy in the eigenvalue:
%
\begin{gather}
\frac{\delta}{\delta \psi^*}\left[
    E_p - \varepsilon \left(  \int d\textbf{r} |\psi(\textbf{r})|^2 - 1  \right)
    \right] = 0 \quad \longrightarrow  \nonumber\\
\label{our_first_polaron_eq_1}
    \left( \frac{1}{2} \sum_{\kappa \kappa^{\prime}} \Phi_{\kappa \kappa^{\prime}} u^0_{\kappa} u^0_{\kappa^{\prime}}
    + \sum_{\kappa} \frac{\partial V_{\textrm{KS}}}{\partial u^0_{\kappa}} u^0_{\kappa} + H_{\textrm{KS}} \right) \psi (\textbf{r})
    = \varepsilon\, \psi(\textbf{r}).
\end{gather}
The solution of this equation is identical to its version without the elastic energy term proposed in Ref.~\cite{Sio201902}. However, now the polaron eigenvalue $\varepsilon$ can be directly interpreted as the polaron energy since from Eqs.~(\ref{our_polaron_energy_functional}) and (\ref{our_first_polaron_eq_1}) one obtains $E_p \left[ \psi, u^0_\kappa \right]=\varepsilon$. Therefore, the polaron formation energy is now given by
\begin{align}
\label{our_formation_energy}
\Delta E_f = \varepsilon - \varepsilon_{\textrm{CBM}},
\end{align}
consistent with Eq.~(\ref{our_formation_energy_1}) in the canonical transformation formalism, because the elastic energy has now been absorbed in the polaron eigenvalue $\varepsilon$.
\\

\newpage
\providecommand{\noopsort}[1]{}\providecommand{\singleletter}[1]{#1}%

\end{document}